# Experiments on Internet Response

Anders Johansen

*Teglgårdsvej 119, DK-3050 Humlebæk, Denmark*

**Abstract.** This paper suggests a generalized distribution of response times to new information $\sim t^{-b}$ for human populations in the absence of deadlines. This has important implications for psychological and social studies as well the study of dynamical networks such as the WWW.

The WWW and Internet email provides for two very efficient methods of distributing information. The advantage compared to old communication channels (Newspapers, TV, ... ) is that both are computer-based and access in principle unrestricted. They thus provide a unique opportunity to study in real time how *fast* humans react to a new piece of information. Along a complementary line of research the WWW has provide a similarly unique opportunity to study evolving networks. Most studies of the WWW have until now focused on the statistical properties, such as the topology. As the WWW is constantly evolving such investigations only deliver a snap-shot of the network. In order to better describe networks between interacting humans one must estimate characteristics of the "nodes", *e.g.*, psychological traits of humans, as one may otherwise be mislead when generalizing from such snap-shot analysis. Hence, the response to both "global events", *e.g.* a news publication, as well as "individual events", *e.g.* receiving an email, must be analysed in order to fully estimate the dynamical features of the network. With respect to commercial exploitation of the Internet, as well as other applications, it is obviously the response of the exposed group to some new piece of information which is of interest and not the topology of the underlying network. The purpose of the present paper is to provide for a characterization of the response of a group of interacting humans in absence of fixed deadlines. Using empirical data of response time distributions of different groups when exposed to a new piece of information we will show that a surprisingly robust dynamical "law" for human response times exist. We will see experiments, where different groups have a similar distribution of individual response times to different kinds of new information. The experiments compliment each other in several aspects. In the first two experiments, a single piece of information is initially presented to a group as a whole. In the third experiment, new information is distributed from one individual to another through email. Thus, in the two first, the new piece of information presented is *the same* for each individual whereas in the third it is different for most individuals. Similarly, we cannot exclude alternative diffusion channels in the first two whereas in the third only a single diffusion channel exists, namely email. In addition, the population in the first two experiments are relatively homogeneous in a socio-economic context, whereas this is not the case in the third experiment. Last, the first two has a similar overall time scale of $\sim 2$ months, whereas in the third experiment it's approximately 1 week. Nevertheless, we get similar results.

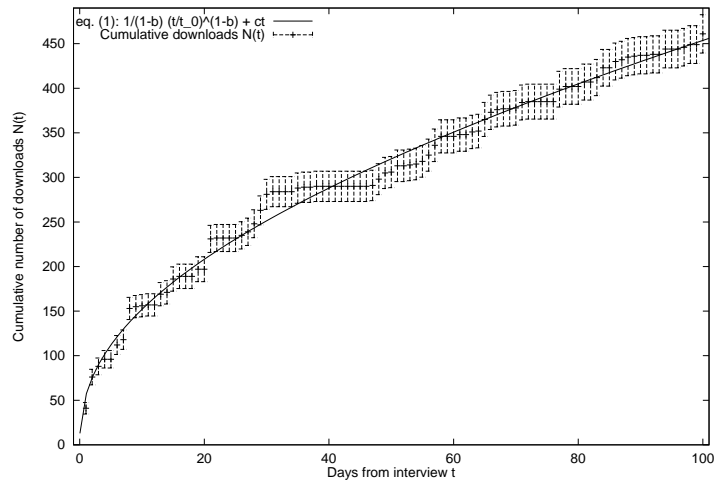

**FIGURE 1.** Cumulative number of downloads as a function of time in days from the appearance of the interview. The fit is $N(t) = a\left(t/t_0\right)^{1-b} + ct$ with $b \approx 0.6$, $t_0 \approx 0.8$ minutes and $c \approx 0.8$ days$^{-1}$

An interview with the author on a subject of broad public interest, namely stock market crashes, was published on the 14th of April 1999 in the business section of one of the leading Danish newspapers. Hence, the population exposed to the new piece of information was largely restricted to regular readers of the business section of the newspaper, which in a socio-economic context is a quite homogeneous population. The text included the URL of the authors homepage with his papers on the subject [1] and it was published both in paper and electronic version on the WWW of the newspaper, the latter restricted to subscribers. The URL to the Los Alamos preprint server, which also contained the author's paper on the subject, was included as well. This means that interested parties had *two options* in order to download papers of interest. In figure 1, we see the cumulative number of downloads $N$ as a function of time $t$ in days from the appearance of the interview. The fit, see caption for details, corresponds to a download rate $n(t) \sim t^{-0.6} + c$. The last term accounts for a constant background taking into account features unrelated to the interview such as search machines and people unaware of the interview. One should note that the wiggles in the data are due to weekends.

The Nasdaq crash culminating on the 14th of April 2000 caught many people with surprise and shook the stock market quite forcefully. As always, many different reasons for the crash were given. The author also had bid for the cause [2], which was made public on the 17th of April 2000 on the Los Alamos preprint server. As a result, a forty minute interview with the author was published on the 26th of May 2000 on a "radio website" together with the URL to the authors papers [1] making it clear that work on stock market crashes in general and the recent Nasdaq crash in particular could be found using the posted URL. The URL of the Los Alamos preprint server was not included on this occasion. Hence a *single* source was provided for the download of the authors papers. Again, it seems reasonable to assume that the population in a socio-economic context is quite homogeneous because of the nature of the Internet-site and the subject. In figure 2, we see the cumulative number of downloads $N(t)$ as a function of time $t$ in days from the appearance of the Web-interview. The fit, see caption for detail, corresponds to a download rate $n(t) \sim t^{-1} + c$. Thus, we again obtain a power law relaxation of the download rate, however, with an exponent $b \sim 1$. The deviation after $\approx 60$ days is due to another advertisement of [1].

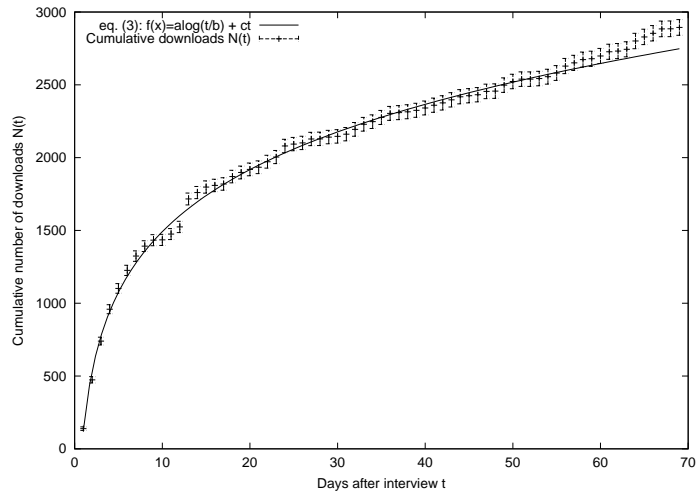

**FIGURE 2.** Cumulative number of downloads as a function of time in days from the appearance of the Web-interview. The fit is $N(t) = a\ln(t/t_0) + ct$ with $t_0 \approx 0.80$ days and $c \approx 2.2$ days$^{-1}$.

In a recent preprint [3], the temporal dynamics of an email network was investigated by J.P. Eckmann, E. Moshes and D. Sergi. Among other things, the time period between the time when an email was sent and its reply received was recorded. The data set contained 3188 users interchanging 309129 emails and was obtained from the log-files from one of the main mail servers of a university. In one sense, this experiment is the "cleanest" as only a single distribution channel exists. However, as each email message contains different information it is also the "dirtiest" with respect to the information distributed. As many different people use the email facility on a university, *e.g*, technical and administrative personnel, students, researchers, ... it seems reasonable to assume that the population in a socio-economic context is quite heterogeneous. In figure 3, we see the cumulative distribution of time periods $\tau$ in hours between the sending of an email messages and its reply. The fit, see caption for details, corresponds to a response time rate $n(\tau) \sim (\tau + c)^{-1}$, over *three decades*. The constant $c$ as a first approximation incorporates the fact that the measured time is not the true response time as most people do not download new emails instantaneously but instead every 10-20 minutes or so. The observant reader will note that the time unit used in this experiment (hours) is a factor 24 smaller than the one used in the previous two. This is presumably due to social conventions, which demand a swifter response to a "private communication" than a "non-private communication", as was the case in the two first experiments. One should note that the wiggles of the data are due to people sharing similar working hours, *i.e.*, emails sent just before leaving work are answered on the following day. The main result to be extracted from this experiment compared to the previous two is that the functional form of the response time distribution of a population exposed to new information is largely independent of the specific nature of the new information.

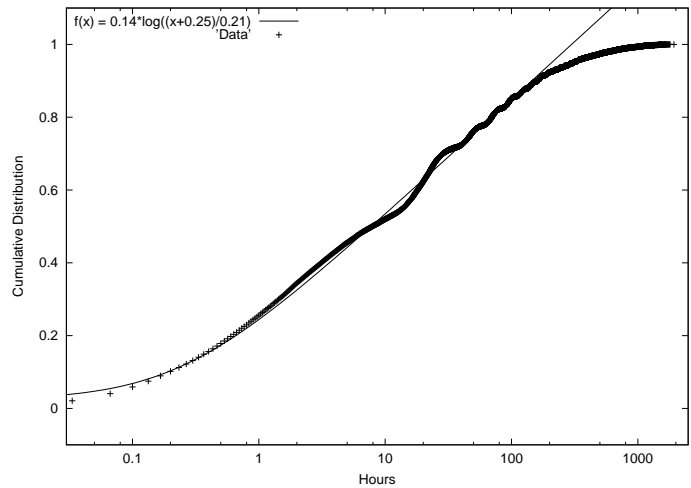

**FIGURE 3.** Cumulative distribution of time periods in hours between an email messages and its reply. The fit is $N(\tau) = a \ln\left((\tau + c)/t_0\right)$ with $t_0 = 4.0$ hours and $c = 0.225$ hours.

In [4] A. G. Chessa and J. M. J. Murre published a cognitive model based on the mathematical theory of point processes to explain an experiment similar to the first two presented here, see their paper for details. In essence, their model is an exponentially decaying function, which the authors claim "proved to give much better fits to our data set and also to the data set described in Johansen", *i.e.*, the data of the second experiment. This is clearly wrong, as can be seen from figures 2 and 4: the distributions shown *does not* correspond to exponential functions. This misinterpretation comes from the fact that Chessa & Murre do not use the cumulative distribution of hits. As "integration" acts as a low-pass filter, it filters out high-frequency noise and is hence much preferable as illustrated here. As the data only contains 28 points, the reader should not too much emphasis on the parameter values obtained from the fit, see caption of figure 4. The prime message of figure 4 is that the cumulative number of hits is very nicely parametrized by a function which yields a similar response time rate as found in the previous 3 experiments and that the value of the exponent is not very different from that obtained from the first experiment of this paper. However, the value $c \approx 0.86$ days is "completely of" and is presumably due to incorrect measurement of starting time $t = 1$ days. I have sent these results to the authors and asked for details on how the time from the appearance of the interview was measured etc. but got no response.

The 3 rather unique experiments presented here shows a self-similar dynamical response of a population to a "perturbation" in the form of new information even though *a priori* the nature of the "perturbations" were quite different. Overall, the experiments suggest that in the absence of formal deadlines the response time rate of human populations to new information to first order have a power law distribution throughout the population. The existence of such a fat-tailed distribution of response times of "agents" has quite important implications for the modeling of social systems such as the financial markets as well as modeling of certain psychological processes. It's quite fascination to realize that similar relaxation behaviour have been observed in such different systems as after-shocks in earthquakes [5], stock markets after crashes [6, 7] and spin glasses [8]. In terms of an reasonable physical analogy, the Trap Model of spin glasses [8] is a personal favourite for the following reasons. At time $t$ after the appearance of the interview, the exposed population consists of two groups, namely those who have not downloaded a paper and those who have. Similarly with respect to the third experiment, at any time $t$ the population considered consists of two groups, namely those who have an email to answer and those who have not. The transition from the first state to the second demands the crossing of some threshold specific to each individual. We thus imagine that the announcement of the URL/the reception of emails plays the role a "field" to which the exposed population is subjected and study the relaxation process by monitoring the number of downloads/the number of replies as a function of time. Hence, we may view the process of downloading/replying as a diffusion process in a random potential, where the act of downloading/replying is similar to that of a barrier-crossing in the Trap Model of spin glasses. These considerations together with the empirical evidence presented strongly suggest that the variation in response time of individuals when confronted with a new piece of information and in the absence of formal deadlines is a power law distribution of waiting times. Each individual thus has a "psycho-sociological barrier" which must be crossed in order for the individual to act and the time needed to cross this barrier is power law distributed throughout the population.

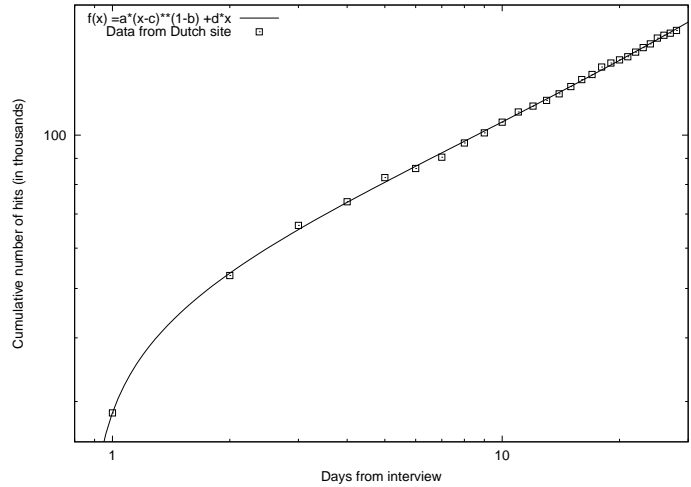

**FIGURE 4.** The cumulative number of hits as a function of time in days from the appearance of the radio broadcast. The fit is $N(t) = a(t-c)^{1-b} + dt$ with $b \approx 0.7$, $c \approx 0.86$ days and $d \approx 1.01 \times 10^3$ days$^{-1}$.

The most pressing question left unanswered by the results presented here is whether each individual has his/her own average characteristic response time and that the power law response time distribution obtained is a result of a population averaging, *i.e.*, the power law distribution obtained is a consequence of differences in each individuals psychology. However, we all know from personal experience that we react with different speed to more or less the same piece of new information depending on personal circumstances at that specific time. In order to answer this question one would ideally like to extract each individuals response time to a number of, for example, emails and this way interchange "ensemble averaging" with "time averaging". Another possibility would be to look at the response time of each individual to a new posting on one of the now very popular dating services on the WWW.